\documentclass[prl,aps,showpacs,floatfix,twocolumn]{revtex4-1}
\usepackage{color}

\usepackage{colordvi}

\usepackage{graphicx,amssymb,epsfig,amsmath}
\usepackage{epstopdf}

\usepackage{amsmath}
\usepackage{bm}
\usepackage{amsfonts}

\begin{document}
\author{Axel U. J. Lode}
\email{axel.lode@unibas.ch} 
\author{Christoph Bruder}
\title{Fragmented superradiance of a Bose-Einstein condensate in an optical cavity}
\affiliation{Department of Physics, University of Basel, Klingelbergstrasse 82, CH-4056 Basel, Switzerland}

\begin{abstract}
  The Dicke model and the superradiance of two-level systems in a
  radiation field have many applications. Recently, a Dicke quantum
  phase transition has been realized with a Bose-Einstein condensate
  in a cavity.  We numerically solve the many-body Schr\"odinger
  equation and study correlations in the ground state of interacting
  bosons in a cavity as a function of the strength of a driving
  laser. Beyond a critical strength, the bosons occupy multiple modes
  macroscopically while remaining superradiant. This fragmented
  superradiance can be detected by analyzing the variance of
  single-shot measurements.
\end{abstract}
\maketitle

The Dicke quantum phase transition from a normal to a superradiant
state is driven by the cooperativity of the emitters in a light
field~\cite{DickeExplained,DickeOrg,Brandes2003,EsslingerRMP}. It has
recently been realized and studied in a number of different systems
like superconducting qubits coupled to
resonators~\cite{squid1,squid2,Viehmann2011,Ciuti2012}, electromagnetically
coupled quantum dots \cite{QD}, the magnetoresistance of organic
light-emitting diodes~\cite{OLED}, a plasma of Helium
atoms~\cite{Plasma}, a superradiant laser~\cite{Laser}, and a
Bose-Einstein condensate in an optical
cavity~\cite{CavityExp0,CavityExp1,CavityExp2,CavityExp3,Chitra2015}.
All of these systems are formed of constituents which are more complex
than the two-level emitters considered in the Dicke model which makes
its successful application to these systems even more remarkable.

The unique experimental control of Bose-Einstein condensates of
ultracold atoms \cite{BEC:Sodium,BEC:Lithium,BEC:Rubidium} has made
them versatile quantum simulators for other systems, like here, the
Dicke
Hamiltonian~\cite{CavityExp0,CavityExp1,CavityExp2,CavityExp3}.
Bose-Einstein condensates or ultracold Fermi gases, however, cannot
generally be described within a two-level framework since they
constitute many-body systems of interacting
atoms~\cite{CB98,BH,exact_F,Axel_exact}. In many-body systems,
correlations arise due to the interactions between the particles and
hence two-level descriptions~\cite{Cavity2level,Cavity2level2} or
mean-field approaches~\cite{CavityMF2,PitaSandro,Pethick}
might fail to accurately describe
them~\cite{exact_F,Axel_exact,Axel_book}. One striking example for a
correlation effect not captured by mean-field methods is the emergence
of fragmentation~\cite{Spekkens,Split,Bader} in interacting
Bose-Einstein condensed systems: the reduced one-body density matrix
starts to have more than one macroscopic eigenvalue.

In this paper, we will focus on a Bose-Einstein condensate in an
optical cavity and show how the fact that the Bose-Einstein condensate
is an interacting many-body system and not just an ensemble of
two-level systems substantially enriches the phase diagram beyond the
Dicke model. Our focus here is on the emergent phases triggered by interactions in a zero-temperature system. The phase diagram of the noninteracting system at finite temperature (the Dicke-Hepp-Lieb phase transition) is discussed in Ref.~\cite{REFEREEA1}.

In the following, we apply a numerical many-body approach for ultracold
atoms in multimodal cavities and demonstrate that their ground state
in a single-mode cavity exhibits correlations that indicate the
fragmentation of the system. For sufficient pump power, the system
enters a state of \textit{fragmented superradiance}. The
pump power needed to trigger the emergence of correlations
in the ground state of the atoms is generally larger than the pump
power necessary to enter the superradiant state.  The existence of
this third phase modifies the phase diagram of ultracold bosons in an
optical cavity and demonstrates the limitations of the mapping of the
system to the Dicke model which exhibits only two phases for bosons in
single-mode cavities.

\begin{figure}
 \includegraphics[width=0.5\textwidth]{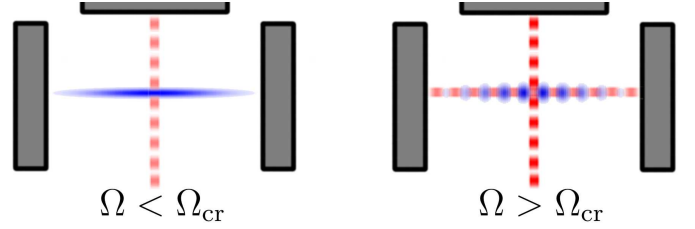}
 \caption{Setup for the Dicke quantum phase transition with a
   Bose-Einstein condensate in an optical cavity. For a cavity pump
   power $\Omega$ that is smaller than the critical pump power
   $\Omega_{\text{cr}}$, the Bose-Einstein condensate (blue) in the cavity formed by the left and right mirrors (gray) is
   unaffected by the pump laser and no population of cavity photons (red) is
   built up (left panel). For pump powers $\Omega$ larger
   than the critical value $\Omega_{\text{cr}}$, the Bose-Einstein
   condensate self-organizes as a consequence of the potential which
   is built up by the population of the cavity with photons (right
   panel). 
}
 \label{Fig:sketch}
 \end{figure}

The realization of the Dicke quantum phase transition with a
Bose-Einstein condensate in a single-mode optical cavity in
\cite{CavityExp0,CavityExp1,CavityExp2,CavityExp3} motivates
us to study the role of correlations in the process of
self-organization. To proceed, we investigate the ground state of a
system of $N=100$ interacting bosonic atoms in a single-mode cavity as
a function of the pump rate, see Fig.~\ref{Fig:sketch} for a scheme of
the system.

The bosons are governed by the time-dependent many-body Schr\"odinger equation  $i \partial_t \vert \Psi \rangle =  H \vert \Psi\rangle$ with the Hamiltonian 
\begin{equation}
 H= \sum_{i=1}^N \left[ T_{\mathbf{r}_i} + V(\mathbf{r}_i) \right] + \lambda_0 \sum_{i>j=1}^N \delta(\mathbf{r}_i, \mathbf{r}_j)\:.
\end{equation} 
Here, $T_{\mathbf{r}_i}$ is the kinetic energy of the $i$-th atom,
$V(\mathbf{r})=V_\text{1-body}(\mathbf{r})+V_\text{cavity}(\mathbf{r})$
is a one-particle potential that contains the external trapping
$V_\text{1-body}$ of the atoms and the potential $V_\text{cavity}$
generated by the photons in the cavity and pump laser beam (see right
panel of Fig.~\ref{Fig:sketch}). The terms proportional to
$\lambda_0$ generate a repulsive contact interaction. We choose
$\lambda_0=0.01$ for the interaction strength $\lambda_0$ which is
proportional to the $s$-wave scattering length and adjustable in
experiment (see Ref.~\cite{units} for experimental parameters to
realize this choice for $\lambda_0$). The potential
$V_\text{cavity}(\mathbf{r})$ is a function of the cavity field amplitude
$\alpha$, which is in turn obtained from the equation of motion of
the cavity \cite{CavityMF2,SUPPLMAT},
\begin{eqnarray}
 i \partial_t \alpha(t) &=& \left[ - \Delta_c + \sum_{k,q=1}^M \rho_{kq}(t) U_{kq}(t) - i \kappa \right] \alpha(t)\nonumber \\ &+& \sum_{k,q=1}^M \rho_{kq}(t) \eta^d_{kq}(t)\:. \label{EOMc} 
\end{eqnarray}
Here, we used
$\langle \Psi(t) \vert U (\mathbf{r}) \vert \Psi(t)\rangle =
\sum_{k,q=1}^M \rho_{kq}(t) U_{kq}(t)$ and
$\langle \Psi(t) \vert \eta (\mathbf{r},t) \vert \Psi(t)\rangle
=\sum_{k,q=1}^M \rho_{kq}(t) \eta_{kq}(t)$, where $U(\mathbf{r})$ and
$\eta(\mathbf{r},t)$ are proportional to the cavity mode and pump
laser profiles, and $\rho_{kq}$ are the matrix elements of the reduced
one-body density matrix that is normalized to $N$, see \cite{SUPPLMAT}. The cavity is detuned by $\Delta_c$ from the atomic resonance and its loss rate is given by $\kappa$.  We employ the
time-dependent multiconfigurational Hartree method for
indistinguishable particles (MCTDH-X) to compute the ground state of
the many-body system coupled to the equation of motion of the cavity
amplitude, see Supplementary Material~\cite{SUPPLMAT} as well as
Refs.~\cite{ultracold,exact_F,SMCTDHB} for details.

In the following we will consider a one-dimensional setup, i.e., a
collinear arrangement of the pump laser and the trapped atoms, and use the
coordinate $x$ instead of $\mathbf{r}$.  We assume the external
confinement to be harmonic, $V_\text{1-body}(x)=\frac{1}{2} x^2$ and
choose dimensionless units~\cite{units} and cavity
parameters~\cite{cavityunits,SUPPLMAT}. The potential exerted on the bosons by
the photons in the pump laser and the cavity
\cite{CavityExp0,SUPPLMAT} is given by
\begin{equation}
 V_\text{cavity}(x)=\vert \alpha \vert^2 U_0 \cos^2(kx) + \left( \alpha+\alpha^*\right) \eta \cos(kx)\:.
\label{cavity_potential}
\end{equation}
Here, the terms proportional to $U_0$ and $\eta$ refer to the cavity
photons and the pump laser, respectively, see Eq.~\eqref{EOMc} and
Supplementary Material \cite{SUPPLMAT}.

Since our system is one-dimensional and parabolically confined, we expect to discover physics different from previous investigations in two-dimensional systems in a lattice, where superfluid self-organized and Mott-insulator self-organized~\cite{REFEREEB1} as well as supersolid and charge-density wave~\cite{REFEREEB2,Hofstetter2013} phases have been demonstrated.

\begin{figure}
 \includegraphics[width=0.5\textwidth]{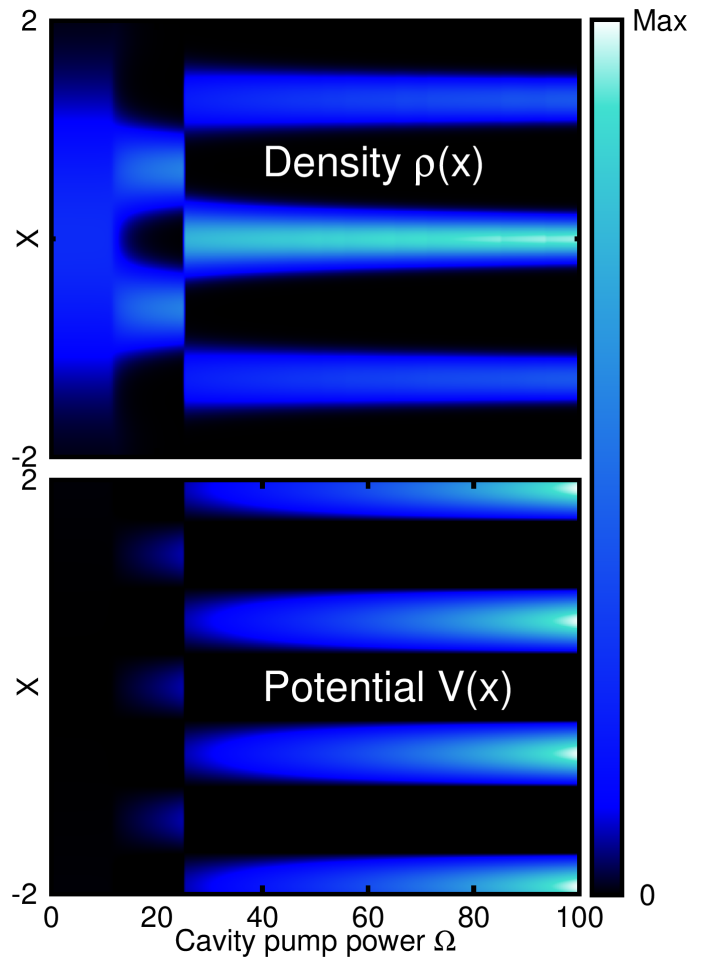}
 \caption{Self-organization of the ground state of a Bose-Einstein
   condensate. The density $\rho(x)$ of the atoms and the potential
   $V(x)=V_\text{1-body}(x) + V_\text{cavity}(x)$ is shown as a
   function of the cavity pump power in the upper and lower panels,
   respectively. Once the applied pump power exceeds a critical
   value, the atoms self-organize because the field which is built up
   inside the cavity creates a periodic one-body potential (cf. lower
   and upper panels). As a result of a competition between external
   and cavity potential as well as interactions, the sign of the
   cavity amplitude switches one time for $\Omega\approx25$
   [compare pattern in density and potential with the inset
   of Fig.~\ref{Fig:frag_pump}(a)].  All quantities shown are
   dimensionless, see text for further discussion.}
 \label{Fig:density}
\end{figure}

\begin{figure}
 \includegraphics[width=0.5\textwidth]{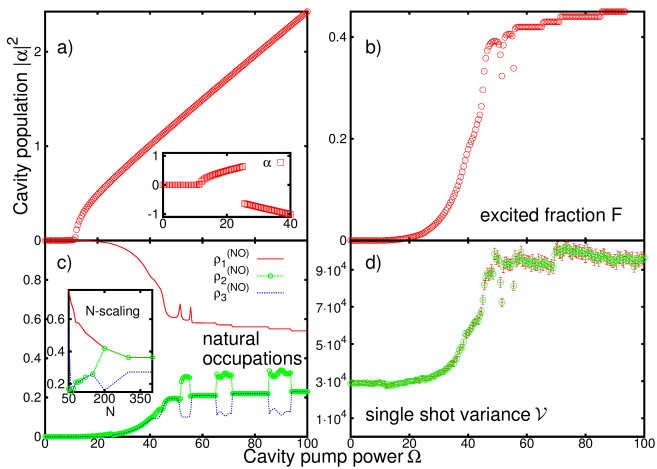}
 \caption{(a) Buildup of cavity population in the self-organization
   process.  Beyond the critical value $\Omega_{\text{cr}}\approx12$
   the population of photons in the cavity grows with increasing pump
   power $\Omega$.  The cavity population $|\alpha|^2$ drives the
   self-organization of the atoms, see
   Eq.~(\ref{cavity_potential}). Inset: The amplitude $\alpha$ shows a sign
   changes in the superradiant phase at
   $\Omega \approx 25$ and is monotonous in the
   fragmented superradiant phase.  (b) Fraction $F$ of atoms that are
   outside the natural orbital with the largest occupation, (c)
   Natural orbital occupations, and (d) Variance $\mathcal{V}$ in
   single shots of the momentum density ($1000$ samples per data
   point), all as a function of the pump power $\Omega$.  The variance
   $\mathcal{V}$ maps the fraction $F$ of atoms in excited orbitals
   closely, compare panels (b) and (d). The natural occupations (c)
   show that the dips in the excited fraction $F$ and the variance are
   due to a re-distribution of atoms between excited orbitals. The inset of (c) demonstrates that fragmented superradiance emerges for a range of particle numbers $N$ with a finite-size scaling for $\Omega=80$. The
   emergence of fragmentation and the growth of $F$ signals that the
   system enters a new phase with many-body correlations between the
   atoms. This is confirmed by comparing (b) and (c) with
   Fig.~\ref{Fig:corr} (a)--(c).  All quantities shown are
   dimensionless, see text.}
 \label{Fig:frag_pump}
\end{figure}

As a first step in our investigation, we show the density $\rho(x)$
and the one-body potential $V(x)$ of the ground state as a function of
the cavity pump power in Fig.~\ref{Fig:density}.
Beyond the critical pump power $\Omega_{\text{cr}}$, the cavity
population $|\alpha|^2$ rises, see Fig.~\ref{Fig:frag_pump}(a), and is
roughly proportional to the magnitude of the maxima of the potential
shown in the lower panel of Fig.~\ref{Fig:density}.
The density becomes \textit{self-organized}
and the atoms cluster around the minima of the cavity-photon-mediated
potential $V_\text{cavity}$ (compare upper and lower panel in
Fig.~\ref{Fig:density}) instead of the minimum of the parabolic
confinement $V_\text{1-body}(x)$. This self-organization of the atoms marks
the transition to the superradiant phase of the Dicke
model~\cite{CavityExp2,CavityExp3}. The emergent density
resembles the density of atoms in optical lattices.

To assess if correlations between the atoms are built up in their
self-organization process, we investigate the cavity population and
fragmentation of the system as a function of the cavity pump power. We
quantify the fragmentation using the fraction $F$ of atoms which do
not occupy the lowest eigenstate of the reduced one-body density
matrix 
\begin{equation}
 \rho^{(1)}(x,x') = \sum_{k} \rho^{(NO)}_{k}
 \phi_k^{(NO),*}(x)\phi_k^{(NO)}(x')\:. 
\label{Eq:rho1}
\end{equation}
The eigenvalues $\rho_k^{(NO)}$ and eigenfunctions $\phi_k^{(NO)}(x)$
of $\rho^{(1)}$ are known as natural occupations and natural orbitals,
respectively [superscript $(NO)$].  If $\rho^{(1)}$ has only a
single macroscopic eigenvalue, $F=0$, the corresponding system is
referred to as condensed. If, on the contrary, $\rho^{(1)}$ has
multiple macroscopic eigenvalues, $F>0$, the corresponding system is
referred to as fragmented~\cite{Spekkens,Bader,Split}. 
Figures~\ref{Fig:frag_pump}(b) and (c)
show a plot of the excited fraction $F$ and the eigenvalues
$\rho_k^{(NO)}$, respectively.  At pump powers
$\Omega \gtrsim 40>\Omega_{\text{cr}}$, i.e., well into the
superradiant regime, the system fragments.  This transition to
fragmentation indicates the emergence of correlations and marks the
break-down of mean-field approaches like the Gross-Pitaevskii
equation, which cannot capture correlations and
fragmentation~\cite{Axel_book,Axel_exact}. Furthermore, the simple two-level description of the superradiant system ceases to be applicable for the fragmented superradiant system, see Sec.~IV of~\cite{SUPPLMAT} for details.

The occurrence of this fragmented superradiant phase is one of the
main results of our Letter. This result is robust against variations in the particle number as demonstrated by a finite-size scaling plot, see inset in Fig.~\ref{Fig:frag_pump}c). We now discuss how the predicted
fragmentation may be detected experimentally.  The emergent phase
cannot be detected in the photonic part of the system alone, see
Fig.~\ref{Fig:frag_pump}(a). Its detection requires the simultaneous
analysis of the wavefunction of the atoms in the cavity and the cavity
photons. We have found that the fragmentation of the Bose-Einstein
condensate can be detected in the variance $\mathcal{V}$ of
single-shot measurements~\cite{SingleShots,SUPPLMAT} of the momentum
distribution, see Fig.~\ref{Fig:frag_pump}(d) for a plot of
$\mathcal{V}$. This variance maps the excited fraction of atoms
accurately and can therefore be used to assess the fragmentation of
the system, compare Figs.~\ref{Fig:frag_pump}(b) and (d).

The similarity of the behavior of the excited fraction $F$ and the
single-shot variance $\mathcal{V}$ may be understood qualitatively:
For a coherent condensate, $F \approx 0$, the variance $\mathcal{V}$
is minimized because all the atoms in the respective single-shot
measurements are picked from the same natural orbital.  For a
fragmented state, $F>0$, the variance of the single shots grows, since
the atoms are picked from a superposition of several mutually
orthonormal natural orbitals $\phi_1^{(NO)}$, $\phi_2^{(NO)}$, \ldots.
The momenta obtained by drawing from this distribution that lives in a
larger space spanned by several orbitals have a wider spread. Hence
the variance $\mathcal{V}$ is larger as compared to the values
obtained by drawing momenta from a single orbital. We expect that the
(momentum-space) variance of single-shot measurements can be used to
quantify fragmentation experimentally also in more general
setups. Since single-shot measurements require only absorption images,
this would mark a clear advantage in comparison to other methods to
determine fragmentation that require the measurement of the
off-diagonal part of the reduced one-body density
matrix~\cite{Spekkens,Split,Bader} or density-density correlations~\cite{Fischer}.

\begin{figure}
 \includegraphics[width=0.5\textwidth]{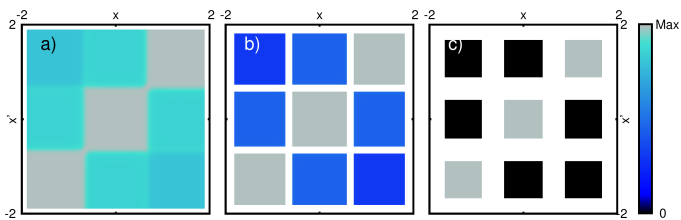}
 \caption{Tracing the transition from superradiance to fragmented
   superradiance in the spatial correlation function. The
   correlation function $\vert g^{(1)}(x,x')\vert^2$ in the
   superradiant phase is shown for pump powers (a) $\Omega=30$,
   (b) $\Omega=40$, and (c) $\Omega=100$, 
   wherever the density exceeds $0.001$. With increasing pump power
   the atoms in distinct wells gradually loose coherence
   ($\vert g^{(1)}(x,x'\neq x)\vert^2$ drops to zero), compare the
   position of black squares with the upper panel of
   Fig.~\ref{Fig:density}. The structure of $\vert g^{(1)}\vert^2$ in
   (c) demonstrates the three-fold fragmentation in the system.  All
   quantities shown are dimensionless.}
 \label{Fig:corr}
\end{figure}

To get a detailed picture of the structure of the emergent many-body
correlations, we analyze the spatial correlation function
$g^{(1)}(x,x')=\frac{\rho^{(1)}(x,x')}{\sqrt{\rho(x)\rho(x')}}$. For a complementary analysis of the momentum
correlation function, see ~\cite{SUPPLMAT}.  The correlation function
$g^{(1)}$ uses the reduced one-body density matrix $\rho^{(1)}$ to
measure the proximity of a many-body state to a product or mean-field
state for a given set of coordinates $x,x'$.  It is a key measure for
the coherence of ultracold bosonic atoms and experimentally detectable
\cite{corr1,corr2}. Figure~\ref{Fig:corr} shows a plot of
$\vert g^{(1)}(x,x')\vert^2$.  The spatial correlation function gives
an intuitive picture of the mechanism behind the fragmentation of the
system. The potential exerted on the atoms by the photons in the
cavity grows with the pump power. Due to the repulsive interparticle
interactions the atoms in the distinct wells of this
cavity-photon-mediated periodic potential become disconnected beyond a
critical magnitude of the pump power.  Consequently, the spatial
correlation function vanishes gradually for off-diagonal values, see
Fig.~\ref{Fig:corr}.  This observation is consistent with the
occurrence of fragmentation, see Figs.~\ref{Fig:frag_pump}(b) and
(c). The loss of spatial coherence between atoms in distinct wells of
the potential generated by the photons in the cavity is complemented
by a periodic pattern of correlated and uncorrelated momenta in the
momentum correlation function that reflects the periodicity of the
potential, see Supplementary Material \cite{SUPPLMAT} and Fig.~S1
therein.

In conclusion, we have derived and applied a new numerical many-body
approach to describe laser-pumped ultracold atoms in cavity fields.
We have used the multiconfigurational time-dependent Hartree method
for indistinguishable particles, MCTDH-X, to solve the many-body
Schr\"odinger equation of the coupled cavity-atom system. This 
method self- and size-consistently incorporates correlations between
the atoms in the cavity and can also be applied to multi-mode cavities
and fermionic atoms.

We have demonstrated that the phase diagram of ultracold bosonic atoms
in a single-mode cavity exhibits a fragmented superradiant phase in
which the atoms show many-body correlations not captured within
two-level or mean-field approaches.  While the system remains
superradiant, the Bose-Einstein condensate starts to macroscopically
occupy multiple single-particle states when the pump power is
increased sufficiently. Thus, our work adds a new phase to the phase
diagram of bosons in a cavity which can be detected using the statistics 
of experimental single shot measurements and the cavity population.

\acknowledgments{We would like to thank E. Fasshauer,
  R. Landig, and M.C. Tsatsos for interesting discussions and in particular R. Chitra, P. Molignini, and L. Papariello for pointing out an inconsistency in the normalization of the reduced one-body density matrix. Financial support by the Swiss SNF and the
  NCCR Quantum Science and Technology is gratefully acknowledged.}


\begin{thebibliography}{References}


\bibitem{DickeExplained} K. Rza\.{z}ewski, K. W\'{o}dkiewicz, and
  W. \.{Z}akowicz, Phys. Rev. Lett. {\bf 35}, 432 (1975).

\bibitem{DickeOrg} R. H. Dicke, Phys. Rev. {\bf 93}, 99 (1954).

\bibitem{Brandes2003} C. Emary and T. Brandes,
Phys. Rev. E {\bf 67}, 066203 (2003).

\bibitem{EsslingerRMP} H. Ritsch, P. Domokos, F. Brennecke, and T. Esslinger,
Rev. Mod. Phys. {\bf 85}, 553 (2013).

\bibitem{squid2} J.~A. Mlynek, A.~A. Abdumalikov, C. Eichler, and
  A. Wallraff, Nature Comm. {\bf 5}, 5186 (2014).

\bibitem{squid1} M. Feng, Y.P. Zhong, T. Liu, L.L. Yan, W.L. Yang,
  J. Twamley, and H. Wang, Nature Comm. {\bf 6}, 7111 (2015).

\bibitem{Viehmann2011} O. Viehmann, J. von Delft, and F. Marquardt,
Phys. Rev. Lett. {\bf 107}, 113602 (2011).
\bibitem{Ciuti2012} C. Ciuti and P. Nataf, Phys. Rev. Lett. {\bf 109}, 179301 (2012).

\bibitem{QD} M. Scheibner, T. Schmidt, L. Worschech, A. Forchel,
  G. Bacher, T. Passow, and D. Hommel, Nature Phys. {\bf 3}, 106 (2007).

\bibitem{OLED} D.~P. Waters, G. Joshi, M. Kavand, M.~E. Limes,
  H. Malissa, P.~L. Burn, J.~M. Lupton, and C. Boehme, Nature
  Phys. {\bf 11}, 910 (2015).

\bibitem{Plasma} H. Xia, A.~A. Svidzinsky, L. Yuan, C. Lu, S. Suckewer, 
  and M.~O. Scully, Phys. Rev. Lett. {\bf 109}, 093604 (2012).

\bibitem{Laser} J.~G. Bohnet, Z. Chen, J.~M. Weiner, D. Meiser,
  M.~J. Holland, and J.~K. Thompson, Nature {\bf 484}, 78 (2012).

\bibitem{CavityExp0} F. Brennecke, R. Mottl, K. Baumann, R. Landig, T. Donner, 
and T. Esslinger, Proc. Natl. Acad. Sci. {\bf 110}, 11763   (2013).

\bibitem{CavityExp1} F. Brennecke, T. Donner, S. Ritter, T. Bourdel,
  M. K\"ohl, and T. Esslinger, Nature {\bf 450}, 268 (2007).

\bibitem{CavityExp2} K. Baumann, C. Guerlin, F. Brennecke, and
  T. Esslinger, Nature {\bf 464}, 1301 (2010).

\bibitem{CavityExp3} J. Klinder, H. Keßler, M. Wolke, L. Mathey,
and A. Hemmerich, Proc. Natl. Acad. Sci. {\bf 112}, 3290 (2015).

\bibitem{Chitra2015} R. Chitra and O. Zilberberg,
Phys. Rev. A {\bf 92}, 023815 (2015).

\bibitem{BEC:Rubidium} M.~H. Anderson, J.~R. Ensher, M.~R. Matthews,
  C.~E. Wiemann, and E.~A. Cornell, Science {\bf 269}, 198 (1995).

\bibitem{BEC:Lithium} C.~C. Bradley, C.~A. Sackett, J.~J. Tollet, and
  R.~G. Hulet, Phys. Rev. Lett. {\bf 75}, 1687 (1995).

\bibitem{BEC:Sodium} K.~B. Davis, M.-O. Mewes, M.~R. Andrews,
  N.~J. van Druten, D.~S. Durfee, D.~M. Kurn, and W. Ketterle,
  Phys. Rev. Lett. {\bf 75}, 3969 (1995).

\bibitem{CB98} D. Jaksch, C. Bruder, J. I. Cirac, C. W. Gardiner, and
  P. Zoller, Phys. Rev. Lett. {\bf 81}, 3108 (1998).

\bibitem{BH} M. Greiner, O. Mandel, T. Esslinger, T. W. H\"ansch, and
  I. Bloch, Nature {\bf 415}, 39 (2002).

\bibitem{exact_F} E. Fasshauer and A.~U.~J. Lode, Phys. Rev. A {\bf
    93}, 033635 (2016).

\bibitem{Axel_exact} A. U. J. Lode, K. Sakmann, O. E. Alon,
  L. S. Cederbaum, and A. I. Streltsov, Phys. Rev. A {\bf 86}, 063606 (2012).

\bibitem{Cavity2level2} D. Nagy, G. K\'{o}nya, G. Szirmai, and
  P. Domokos, Phys. Rev. Lett. {\bf 104}, 130401 (2010).

\bibitem{Cavity2level} M. Paternostro, G. De Chiara, and G. M. Palma,
  Phys. Rev. Lett. {\bf 104}, 243602 (2010).

\bibitem{Pethick} C.\,J. Pethick and H. Smith, {\em Bose-Einstein
    Condensation in Dilute Gases} (Cambridge University Press,
  Cambridge UK, 2002).

\bibitem{PitaSandro} L.\,P. Pitaevskii and S. Stringari, {\em
    Bose-Einstein Condensation} (Clarendon Press, Oxford, 2003).

\bibitem{CavityMF2} D. Nagy, G. Szirmai, and P. Domokos,
Eur. Phys. J. D {\bf 48}, 127 (2008).

\bibitem{Axel_book} A.\,U.\,J. Lode, {\em Tunneling Dynamics in Open
    Ultracold Bosonic Systems}, Springer Theses, Springer, Heidelberg
  (2014).

\bibitem{Spekkens} R.\,W. Spekkens and J.\,E. Sipe, 
   Phys. Rev. A {\bf 59}, 3868 (1999).

\bibitem{Split} A.~I. Streltsov, O.~E. Alon, and L.~S. Cederbaum,
  Phys. Rev. Lett. {\bf 99}, 030402 (2007).

\bibitem{Bader} P. Bader and U.\,R. Fischer,
Phys. Rev. Lett. {\bf 103}, 060402 (2009). 

\bibitem{REFEREEA1} F. Piazza, P. Strack, W. Zwerger, Annals of Physics {\bf 339}, 135 (2013).

\bibitem{units}  We first choose a length scale of $L=1\mu$m. The
  scale of energy for the mass of $^{87}$Rb is
  $\hbar^2/(m L^2)=2\pi\hbar\times 116$Hz, and the scale of time is
  $mL^2/\hbar=1.37$ms. The one-dimensional scattering parameter
  $\lambda_0$ is related to the three-dimensional scattering length
  $a_{3D}$ by $\lambda_0=2Lm\omega_\perp a_{3D}/\hbar$ where
  $\omega_\perp$ is the frequency of the transversal
  confinement~\cite{Olshanii}.  Using $a_{3D}=100.4 a_0$, where $a_0$
  is the Bohr radius, and $\lambda_0=0.01$, one obtains
  $\omega_\perp =687.9$Hz.

\bibitem{Olshanii} M. Olshanii, Phys. Rev. Lett. {\bf 81}, 938 (1998).

\bibitem{SUPPLMAT} See supplementary material.drawn 

\bibitem{invariance} H.-D. Meyer, U. Manthe, and L.~S. Cederbaum, 
Chem. Phys. Lett. {\bf 165}, 73 (1990); U. Manthe, H.-D. Meyer, and L.~S. Cederbaum, 
J. Chem. Phys. {\bf 97}, 3199 (1992).

\bibitem{alon} O.~E. Alon, A.~I. Streltsov, and L.~S. Cederbaum, 
Phys. Rev. A {\bf{77}}, 033613 (2008).
\bibitem{MCTDHX} O.~E. Alon, A.~I. Streltsov, and L.~S. Cederbaum, 
J. Chem. Phys. {\bf 127}, 154103 (2007).
\bibitem{Kristian} K. Baumann, {\it Experimental Realization of the Dicke
  Quantum Phase Transition}, Ph.D. Thesis No. 19943, ETH Z\"urich (2012).
\bibitem{Penrose} O. Penrose and L. Onsager, 
Phys. Rev. \textbf{104}, 576 (1956).

\bibitem{SMCTDHB} A.U.J. Lode, Phys. Rev. A {\bf 93}, 063601 (2016).
\bibitem{ultracold} A.\,U.\,J. Lode, M.\,C. Tsatsos, and E. Fasshauer,
  \textsc{MCTDH-X}: {\em The time-dependent multiconfigurational
    Hartree method for indistinguishable particles software},
  \href{http://ultracold.org}{http://ultracold.org}.

\bibitem{cavityunits} In the units of Ref.~\cite{units}, the cavity
  detuning is $\Delta_c=43$, the cavity loss rate $\kappa=11.2$, and
  the cavity-atom coupling $g_0=46$. The cavity and pump mode
  functions are given by $\cos(k_c x)$, where the $k_c$-value of the cavity
  is $k_c=4.9$.

 
\bibitem{REFEREEB1} M. Reza Bakhtiari, A. Hemmerich, H. Ritsch, and M. Thorwart, Phys. Rev. Lett. {\bf 114}, 123601 (2015).
\bibitem{REFEREEB2}  R. Landig, L. Hruby, N. Dogra, M. Landini, R. Mottl, T. Donner, and T. Esslinger, Nature {\bf 532}, 476 (2016).
\bibitem{Hofstetter2013} Y. Li, L. He, and W. Hofstetter, Phys. Rev. A {\bf 87}, 051604 (2013).  
  
\bibitem{SingleShots} K. Sakmann and M. Kasevich, Nature Phys. {\bf 12}, 451 (2016).

\bibitem{Fischer} M.~K. Kang and U.~R. Fischer, 
Phys. Rev. Lett. {\bf 113}, 140404 (2014).

\bibitem{corr1} T. Betz, S. Manz, R. B\"ucker, T. Berrada, Ch. Koller,
  G. Kazakov, I.E. Mazets, H.-P. Stimming, A. Perrin, T. Schumm, and
  J. Schmiedmayer,
Phys. Rev. Lett. {\bf 106}, 020407 (2011). 

\bibitem{corr2} A. Perrin, R. B\"ucker, S. Manz, T. Betz, C. Koller,
 T. Plisson, T. Schumm, and  J. Schmiedmayer,
Nature Phys. {\bf 8}, 195 (2012). 


\end{thebibliography}
\end{document}


\author{Axel U. J. Lode}
\email{axel.lode@unibas.ch} 
\author{Christoph Bruder}
\title{Supplementary Information\\Fragmented superradiance of a Bose-Einstein condensate in an optical cavity}
\affiliation{Department of Physics, University of Basel, Klingelbergstrasse 82, CH-4056 Basel, Switzerland}

\maketitle

The results of the paper are based on the numerical method described
in Sec.~\ref{Sec:methods} of this supplementary document. In
Sec.~\ref{Sec:Shots} we discuss the algorithm used to simulate
experimental single-shot measurements, in Sec.~\ref{Sec:Corr} we
present calculations of the momentum correlation function that gives
additional insights on the build-up of correlations and fragmentation of the condensate, and in Sec.~\ref{Sec:2level} we discuss the breakdown of the two-level description in the fragmented superradiant phase.

\section{Numerical Method}\label{Sec:methods}
\subsection{Hamiltonian}
Our goal is to solve the time-dependent many-body Schr\"odinger equation,
\begin{equation}
 i \partial_t \vert \Psi \rangle =  \hat{H} \vert \Psi\rangle\:,
\label{TDSE}
\end{equation}
where $\hat{H}$ is a general Hamiltonian for a system of ultracold atoms which is pumped by a laser and resides in an optical cavity,
\begin{equation}
 \hat{H}= \sum_{i=1}^N \hat{h} (\vec{r}_i;t) + \sum_{i<j=1}^N
 \hat{W}(\vec{r}_i, \vec{r}_j;t)\:. 
\label{H1}
\end{equation}
Here, $\hat{h}$ is a one-body operator and $\hat{W}$ a two-body
interparticle interaction. The potential felt by the atoms created by
the cavity photons can be cast in the form of a time-dependent
one-body potential $V_\text{cavity}(\vec{r};t)$ and is hence absorbed
in the one-body Hamiltonian $\hat{h}$ in Eq.~\eqref{H1}. To compute
this cavity-photon potential the equation of motion of the cavity (see
below) has to be solved 
simultaneously and coupled to the many-body problem in Eq.~\eqref{TDSE}.

\subsection{MCTDH-X}
We use the multiconfigurational time-dependent Hartree method for indistinguishable particles (MCTDH-X) to compute the time-evolution for a general many-body wavefunction 
\begin{equation}
\vert \Psi (t) \rangle = \sum_{\vec{n}} C_{\vec{n}}(t) \vert \vec{n} ;
t \rangle\:.  
\label{ansatz}
\end{equation}
The basis $ \vert \vec{n};t \rangle$ consists of all possible (anti-)symmetrized products of $N$ particles in $M$ single-particle functions $\lbrace \Phi_i(\vec{r};t),i=1,...,M\rbrace$.
This amounts to a description of the atoms by the field operator $\hat{\Psi}(\vec{r};t) = \sum_{i=1}^M \hat{b}_i(t) \Phi_i^*(\vec{r};t)$. The configurations $\vert \vec{n}; t \rangle$ read 
\begin{equation}
\vert \vec{n} ; t \rangle = \prod_{i=1}^M \left[ \frac{\left(
      \hat{b}^\dagger_i(t) \right)^{n_i}}{ \sqrt{n_i!}} \right] \vert
vac \rangle\:. 
\label{basis}  
\end{equation}

Applying the time-dependent variational principle to the time-dependent Schr\"odinger equation with this ansatz while exploiting an invariance property \cite{invariance} yields the equations of motion of the MCTDH-X \cite{MCTDHX,alon} method: 
\begin{eqnarray} 
i \partial_t \vert \Phi_j \rangle &=& \hat{P} \left[ \hat{h} \vert \Phi_j \rangle + \lambda_0 \sum_{k,s,q,l=1}^{M} \lbrace \boldsymbol{\rho}\rbrace^{-1}_{jk} \rho_{ksql} \hat{W}_{sl}\vert \Phi_q \rangle \right], \label{EOMa} \\
 i \partial_t \boldsymbol{\mathcal{C}}(t) &=& \boldsymbol{\mathcal{H}}(t) \boldsymbol{\mathcal{C}}(t), \label{EOM}
\end{eqnarray}
Eq.~\eqref{EOMa} for the orbitals and Eq.~\eqref{EOM} for the coefficients.
Here, $\rho_{kq}=\langle \Psi \vert \hat{b}^\dagger_k \hat{b}_q \vert \Psi \rangle$ and $\rho_{kqsl}=\langle \Psi \vert \hat{b}^\dagger_k \hat{b}^\dagger_s \hat{b}_l \hat{b}_q \vert \Psi \rangle$, are the matrix elements of the reduced one-body and two-body density matrix, respectively. The local interaction potentials $\hat{W}_{sl}(\vec{r})=\int d\vec{r}' \Phi^*_s(\vec{r}',t) \hat{W}(\vec{r},\vec{r}',t) \Phi_l(\vec{r}',t)$ were introduced, $\boldsymbol{\mathcal{H}}_{\vec{n},\vec{n}'} = \langle \vec{n}' ; t \vert \hat{H} \vert \vec{n};t \rangle$ is the Hamiltonian represented in the chosen many-body basis, and $\boldsymbol{\mathcal{C}}(t)$ is a vector collecting all coefficients $\lbrace C_{\vec{n}}(t) \rbrace$. For the details of the derivation of Eqs.~\eqref{EOMa} and \eqref{EOM}, see Refs.~\cite{MCTDHX,alon}.

\subsection{Coupling a many-body system to a multi-mode cavity}
In the MCTDH-X equations of motion (Eqs.~\eqref{EOMa} and
\eqref{EOM}), the potential exerted on the atoms by the photons
in the cavity and the pump was absorbed into the one-body Hamiltonian $\hat{h}$
which retains its form, $\hat{h}= \hat{T}_{\vec{r}} + V(\vec{r};t)$. As mentioned, $\hat{T}_{\vec{r}}$ is the kinetic energy and $V=V_\text{cavity}(\vec{r};t)+V_\text{1-body}(\vec{r};t)$ the combined external
potential of the trapping potential $V_\text{1-body}$, the pump laser, and the cavity
photons. The potential $V(\vec{r},t)$ is hence modified in a time-dependent
way by the interaction of the atoms with the cavity field
$\alpha^d(t)$. Here and in the following, the superscript $d$ indexes
the cavity modes. The effect of this interaction is two-fold
\cite{Kristian,CavityExp2,CavityExp3}: the field of the cavity generates a
potential $U^d(\vec{r}) = \frac{(g^d_0)^2}{\Delta_a} \left(\xi^d_C(\vec{r})\right)^2$
that contributes to the one-body potential experienced by the atoms
and, second, the atoms scatter photons from the pump into the cavity
field resulting in a cavity photon source term $\eta^d(\vec{r},t) = \frac{g^d_0 \Omega(t)}{\Delta_a} \xi_P(\vec{r})
\xi_C^d(\vec{r})$. Here, $\xi_C^d(\vec{r})$ is the profile of the $d$-th cavity mode, $\xi_P(\vec{r})$ the profile of the pump, $g^d_0$ is the atom-cavity coupling and
$\Delta_a$ is the detuning of the pump laser from resonance frequency
of the atoms. The pump field $V_0(\vec{r};t)=\frac{\Omega^2(t)}{\Delta^2_a} \xi^2_P(\vec{r})$ creates an additional potential for the atoms. Here, the pump Rabi frequency $\Omega$ depends on the pump laser power which may be time-dependent. The resulting one-body potential entering the many-body Hamiltonian of bosons in an optical cavity reads
\begin{equation}
  V(\vec{r},t) =  V_\text{1-body}(\vec{r},t) + V_0(\vec{r};t) + \sum_d \Bigg[ \vert \alpha^d (t) \vert^2 U^d(\vec{r}) + (\alpha^d(t)+\alpha^{*,d}(t)) \eta^d(\vec{r},t)\Bigg]\:, \nonumber
\end{equation}
where $V_\text{1-body}(\vec{r},t)$ is the external trapping potential of the system without the cavity and $\xi_P(\vec{r})$ and $\xi^d_C(\vec{r})$ are the pump beam and cavity profiles, respectively. In the one-dimensional setup considered in the main text, the system resides in a node of the pump and therefore there is no contribution of $V_0$ to the potential. We use the MCTDH-X orbitals (cf. Eq.~\eqref{EOMa}) to define the matrix elements $\langle \Phi_k \vert U^d(\vec{r};t) \vert \Phi_q \rangle = U^d_{kq}(t)$ and  $\langle \Phi_k \vert \eta^d(\vec{r},t) \vert \Phi_q \rangle = \eta^d_{kq}(t)$, i.e., the expectation values which are necessary to evaluate the equation of motion of the cavity field below. 
Instead of solving the Master equation that describes photon loss from the cavity, we introduce a complex damping term $-i\kappa^d$ which allows to cast the problem in the form of a set of coupled equations of motion of the cavity populations~\cite{CavityMF2,Kristian}:
\begin{eqnarray}
i \partial_t \alpha^d(t) &=& \left[ - \Delta^d_c +  \sum_{k,q=1}^M \left( \rho_{kq}(t) U^d_{kq}(t) \right) - i \kappa^d + \sum_{d'} \gamma_{dd'} \alpha^{d',*}(t) \right] \alpha^d(t) \nonumber \\  &+& \sum_{k,q=1}^M \left(\rho_{kq}(t) \eta^d_{kq}(t) \right) + \sum_{kq}\Bigg[ \rho_{kq} \sum_{d'\neq d} \Gamma_{kq}^{dd'}\alpha^{d}(t)\alpha^{d',*}(t) \Bigg]. \qquad  \label{EOMc} 
\end{eqnarray}
Here, we used the relations $\langle \Psi(t) \vert U^d (\vec{r}) \vert \Psi(t)\rangle = \sum_{k,q=1}^M \left( \rho_{kq}(t) U^d_{kq}(t) \right)$ and $\langle \Psi(t) \vert \eta^d (\vec{r},t) \vert \Psi(t)\rangle =\sum_{k,q=1}^M \left(\rho_{kq}(t) \eta^d_{kq}(t)\right)$. The atom-mediated cavity mode coupling between modes $d$ and $d'$ is given by $\Gamma_{kq}^{dd'}=\nu^{dd'}\langle \phi_k \vert \xi_C^d(\vec{r},t) \xi_C^{d'}(\vec{r},t) \vert \phi_q \rangle$ with mode coupling constants $\nu^{dd'}$ while the static cavity-mode coupling of mode $d$ and $d'$ is mediated by the coefficients $\gamma_{dd'}$. 

Equations \eqref{EOMa}, \eqref{EOM}, and \eqref{EOMc} are solved simultaneously to describe the many-body cavity dynamics with MCTDH-X.

Note, that the main text deals with a single-mode cavity for which $\gamma_{dd'}=\Gamma^{dd'}_{kq}=0$ in Eq.~\eqref{EOMc}. For the simulations presented in the main text, we used the cavity detuning $\Delta_C=43$, the cavity loss-rate $\kappa=11.2$, the cavity-atom coupling $g_0=0.46$. The cavity and pump mode functions are given by $\xi_C(x)=\cos(kx)=\xi_P(x)$, where $k=4.9$. All the cavity parameters are given in the dimensionless units defined in Ref.~[34] of the main text.

\subsection{Numerical details} 
We used $M=3$ orbitals $\Phi_k(\vec{r};t)$ to represent the many-body state $\vert \Psi \rangle=\sum_{\vec{n}}C_{\vec{n}} \vert \vec{n}; t \rangle$. With $N=100$, this corresponds to $\binom{N+M-1}{N}=5005$ coefficients $C_{\vec{n}}(t)$. To represent the orbitals, a discrete-variable representation with $1024$ plane-wave functions was employed. The grid ranged from $-8$ to $8$.

\section{Measurement of fragmented superradiance with single shots}\label{Sec:Shots}
Here, we show how to simulate experimental single shot measurements from a many-body wavefunction of Bose-Einstein condensed atoms in an optical cavity. To this end, we describe the algorithm to simulate single shots of the many-body wavefunctions (see Ref.~\cite{SingleShots} for details) which are the result of computations with MCTDH-X~\cite{ultracold}. 

\subsection{Simulation of single shots from many-body wavefunctions}
We follow the procedure developed in Ref.~\cite{SingleShots} to compute single shots 
that can be applied both in real ($\vec{r}$-) and momentum ($\vec{k}$-) space. In the following, we present the equations for real space and use $\vec{r}$ for coordinates.
To start, we define the $N$-body wavefunction
\begin{equation}
 \vert \Psi \rangle = \sum_{\vec{n}} C_{\vec{n}} \vert \vec{n} \rangle.\label{psi}
\end{equation}
Here, the index $\vec{n}$ runs through all possible configurations of $N$ bosons in $M$ single-particle states. Consequently the configurations read $\vert n_1, n_2,...,n_M \rangle$ with $\sum_k n_k =N$. To simulate a single-shot measurement, we define the probability distribution
\begin{equation}
 P(\vec{r}_1,\vec{r}_2,...,\vec{r}_N)=\vert \Psi(\vec{r}_1,\vec{r}_2,...,\vec{r}_N) \vert^2 \label{NProb}
\end{equation}
and note that it can be written as a product of conditional probabilities,
\begin{equation}
 P(\vec{r}_1,\vec{r}_2,...,\vec{r}_N)= P(\vec{r}_1) P(\vec{r}_2 \vert \vec{r}_1) \cdots P(\vec{r}_N  \vert \vec{r}_{N-1},...,\vec{r}_1). \label{CondNProb}
\end{equation}
Here, $P(\vec{r}_a \vert \vec{r}_b, \vec{r}_c )$ represents the
conditional probability to find a particle at $\vec{r}_a$, given that
other particles are found at $\vec{r}_b$ and $\vec{r}_c$, and
analogously for the higher-order conditional probabilities.
A single-shot measurement means to draw the positions of the $N$ particles in the system from the probability distribution given in Eq.~\eqref{NProb}, or, equivalently, from Eq.~\eqref{CondNProb}. In the following we will use Eq.~\eqref{CondNProb}, because its representation requires drawing positions from single-variable conditional distributions and is therefore numerically much less demanding than drawing from the $N$-variable distribution in Eq.~\eqref{NProb}. To calculate the conditional probabilities $P(\cdot \vert \cdot)$, we introduce the \textit{reduced} states
\begin{equation}
 \vert \Psi^d \rangle = \begin{cases} 
                          \vert \Psi \rangle & \text{for } d=0 \\ \\
                          \mathcal{N}_d \hat{\Psi}(\vec{r}_d) \vert \Psi^{d-1} \rangle & \text{for } 0<d<N
                        \end{cases}\label{redPSI}
\end{equation}
Here, $\vec{r}_d$ is the position of the drawn particle, and $\mathcal{N}_d$
is a normalization factor. The reduced $(N-d)$ and $(N-d-1)$-particle
states for a general many-body state $\vert \Psi \rangle$ of the form
in Eq.~\eqref{psi} read $\vert \Psi^d \rangle = \sum_{\vec{n}}
C_{\vec{n}}^d \vert \vec{n} \rangle$ and $\vert \Psi^{d-1} \rangle =
\sum_{\vec{n}} C^{d-1}_{\vec{n}} \vert \vec{n} \rangle$,
respectively. The expansion coefficients obey
\begin{equation}
 C_{\vec{n}}^d = \mathcal{N}_d \sum_{k=1}^M \Phi_k(\vec{r}_d) C_{\vec{n}^k}^{d-1} \sqrt{n_k+1}.\label{redcoeff}
\end{equation}
Here, $\Phi_k(\vec{r}_d)$ are the single-particle wave functions
introduced in Eq.~\eqref{EOMa}, $\vec{r}_d$ is the position of the drawn particle, and $\vec{n}^k=(n_1,...,n_k+1,...,n_M)$ was introduced.
In the following, we will use the reduced wavefunction $\vert \Psi^d \rangle$, that describes $N-d$ particles to evaluate the conditional probabilities in Eq.~\eqref{CondNProb}. For this purpose, we define the one-body densities that correspond to the reduced wavefunctions:
\begin{equation}
 \rho_d(\vec{r}) = \langle \Psi^d \vert \hat{\Psi}^\dagger(\vec{r}) \hat{\Psi}(\vec{r}) \vert \Psi^d \rangle\label{redDENS}
\end{equation}
The normalization $\mathcal{N}_d$ in Eq.~\eqref{redPSI} is hence $\mathcal{N}_d = \frac{1}{\sqrt{\rho_{d-1} (\vec{r}_d)}}$. 
To obtain a single-shot measurement, i.e., the positions of all particles $\vec{r}_1,...,\vec{r}_N$, the first position $\vec{r}_1$ is drawn from $P(\vec{r})=\rho_0(\vec{r})/N$, the second position $\vec{r}_2$ is drawn from $P(\vec{r}\vert \vec{r}_1)$, etc.. To obtain the conditional probabilities $P(\vec{r} \vert \cdot)$, we note that $P(\vec{r} \vert \vec{r}_1,\vec{r}_2,...,\vec{r}_d) =\frac{P(\vec{r}, \vec{r}_1,\vec{r}_2,...,\vec{r}_d)}{P(\vec{r}_d,...,\vec{r}_1)}$ (cf. Eq.~\eqref{CondNProb} and Eq.~\eqref{redPSI}). Furthermore, note that $P(\vec{r}_d,...,\vec{r}_1)$ is a constant. Hence, the conditional probability $P(\vec{r} \vert \vec{r}_d,...,\vec{r}_1)$ is proportional to the density $\rho_d(\vec{r})$.

To summarize, a single-shot measurement can be simulated as follows: Start by drawing a particle position $\vec{r}_d$ from $\rho_{d-1}(\vec{r})$. Then compute $\Psi^d$ from $\Psi^{d-1}$ using Eqs.~\eqref{redPSI} and \eqref{redcoeff}. By normalization, $\mathcal{N}_d$ is determined. From $\Psi^d$, the conditional probability $P(\vec{r} \vert \vec{r}_d,...,\vec{r}_1)$ is evaluated by computing the density $\rho_d(\vec{r})$, see Eq.~\eqref{redDENS}. The next particle position $\vec{r}_{d+1}$ is now drawn from this conditional probability and so forth. These steps are repeated for $d=1,...,N-1$ until all particle positions $\vec{r}_1,...,\vec{r}_N$ have been drawn. To simulate a single-shot measurement, $\mathcal{B}(\vec{r})$, the positions $\vec{r}_1,...,\vec{r}_N$ are then convoluted with a point spread function. Following Ref.~\cite{SingleShots}, we use a Gaussian with a width of three pixels.

\subsection{Fragmented superradiance and the variance of single shots of the momentum distribution}
Here, we derive a quantity to assess fragmentation from experimental single-shot measurements. We use the above algorithm to simulate experimental single shots to obtain the variance of single-shot measurements in the phase transition to fragmented superradiance discussed in the main text. Since a one-dimensional system is considered here, we will use $k$ synonymously to $\vec{k}$ in the following. We define the variance of a set of single-shot measurements $\lbrace \mathcal{B}_j (k) \rbrace_{j=1}^{N_\text{shots}}$ as follows
\begin{equation}
  \mathcal{V}= \int dk \frac{1}{N_\text{shots}} \sum_{j=1}^{N_\text{shots}} \left[ \mathcal{B}_j(k) - \bar{\mathcal{B}}(k) \right]^2;\qquad \bar{\mathcal{B}}(k) = \frac{1}{N_\text{shots}} \sum_{j=1}^{N_\text{shots}} \mathcal{B}_j(k)\:.
\end{equation}
We plot the momentum space single-shot variance $\mathcal{V}$ in Fig.~4 in the main text. 
The variance clearly maps the behavior of the fragmentation and may
hence be used as a straightforward way to experimentally detect the
fragmented superradiant phase from time-of-flight images. The similarity of the
behavior of fragmentation $F$ and single-shot variance $\mathcal{V}$
may be understood qualitatively: For a coherent state, $F\approx0$, the variance $\mathcal{V}$ is minimized because all the particles in the respective single-shot measurements are picked from the same single-particle state $\phi$. For a fragmented state, $F>0$, the variance of the single shots grows, because the particles are picked from a superposition of several mutually orthonormal single-particle states $\phi_1,\phi_2,...$. The momenta obtained by drawing from this distribution that lives in a larger space spanned by several functions have a wider spread, and hence larger variance $\mathcal{V}$ as compared to the values obtained by drawing momenta from a single function. We speculate that the (momentum space) variance of single shots can be used to quantify fragmentation experimentally in general setups and not only the one considered here and in the main text. Since single shots require only absorption images, this would mark a clear advantage in comparison to other methods to determine fragmentation 
that 
necessitate the 
measurement of correlations~\cite{Fish} or the off-diagonal part of the reduced one-body density matrix~\cite{Penrose}.

\section{Tracing the fragmented superradiant phase in the momentum space correlations of the atoms}\label{Sec:Corr}
\begin{figure}
 \includegraphics[width=\textwidth]{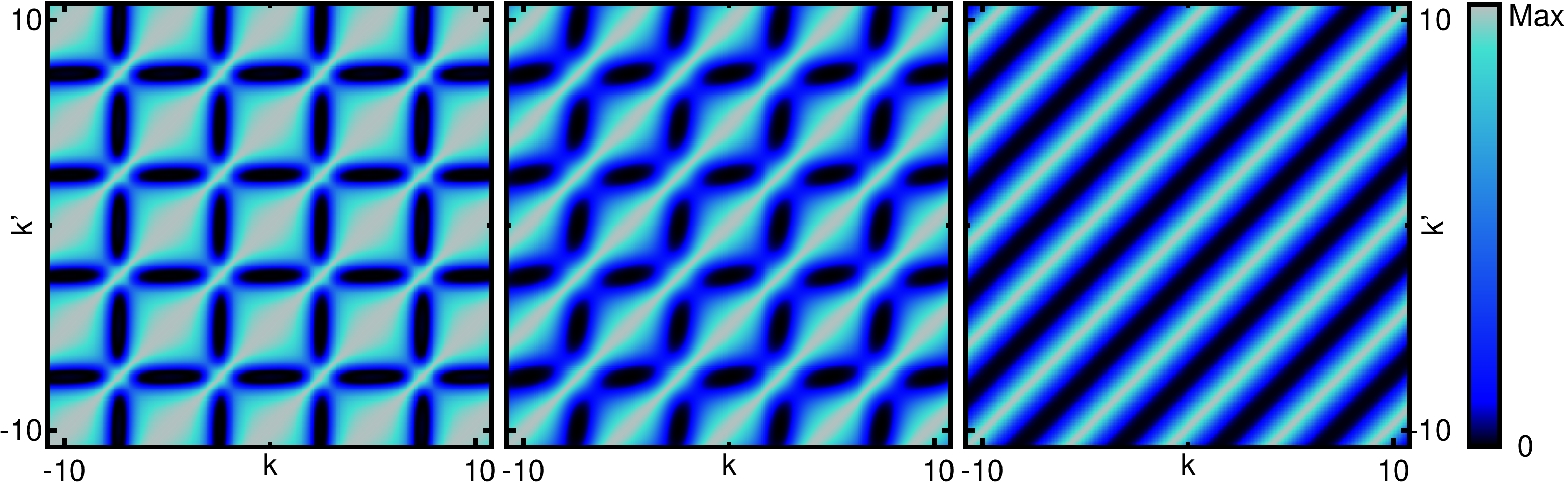}
 \caption{Tracing the transition from superradiance to fragmented superradiance in the momentum correlation functions of the atoms. The momentum correlation functions $\vert g^{(1)} (k,k') \vert^2$ in the superradiant phase are shown for pump-powers $\Omega=30,40,100$ in the left, middle, and right panels, respectively, wherever the momentum density exceeds $0.001$. 
  Overall, the momentum correlations are more sensitive to fragmentation than their real-space counterpart (cf. Fig.~4 in the main text): even at the lowest depicted pump rate, $\Omega=30$, the momentum correlations show a square 
lattice pattern where $\vert g^{(1)}\vert^2<1$, 
although the sample is still almost fully coherent (no fragmentation, $F\approx 0$). Through the transition, the square lattice pattern transforms into a diagonal line 
pattern of alternating domains of uncorrelated ($\vert g^{(1)} \vert^2 \approx 1$) and correlated momenta ($\vert g^{(1)} \vert^2 \approx 0$, cf. middle and right panels). All quantities shown are dimensionless.}
 \label{Fig:corr}
\end{figure}

We now consider the momentum correlations 
during the transition from superradiance to fragmented superradiance using
$g^{(1)}(k,k') = \frac{\rho^{(1)}(k,k')}{\sqrt{\rho(k)\rho(k')}}$,
here $\rho^{(1)}(k,k')$ is the reduced one-body density matrix in
momentum space. It can be obtained by applying a Fourier transform to
the natural orbitals $\phi_k^{(NO)}(x)$ in Eq.~(4) in the main
text. Figure~\ref{Fig:corr} shows the momentum correlations
$|g^{(1)}(k,k')|^2$ as a function of the pump power $\Omega$.
The momentum correlations appear to be more susceptible to
fragmentation, as they show the loss of coherence already for very
small values of the excited fraction $F$, compare Fig.~\ref{Fig:corr}
and Fig.~3(b) of the main text. The pattern of squares that
emerges in the transition to fragmented superradiance in the spatial
correlations [cf. Fig.~4(a)--(c) of the main text] is visible in the
momentum correlations as a periodic structure of diagonal lines of
alternating coherent ($\vert g^{(1)}\vert^2 \approx 1$) and incoherent
($\vert g^{(1)} \vert^2 \approx 0$) momenta, see right panel in
Fig.~\ref{Fig:corr}.

\section{Breakdown of the two-level description in momentum space}\label{Sec:2level}

\begin{figure}[!]
 \includegraphics[width=\textwidth]{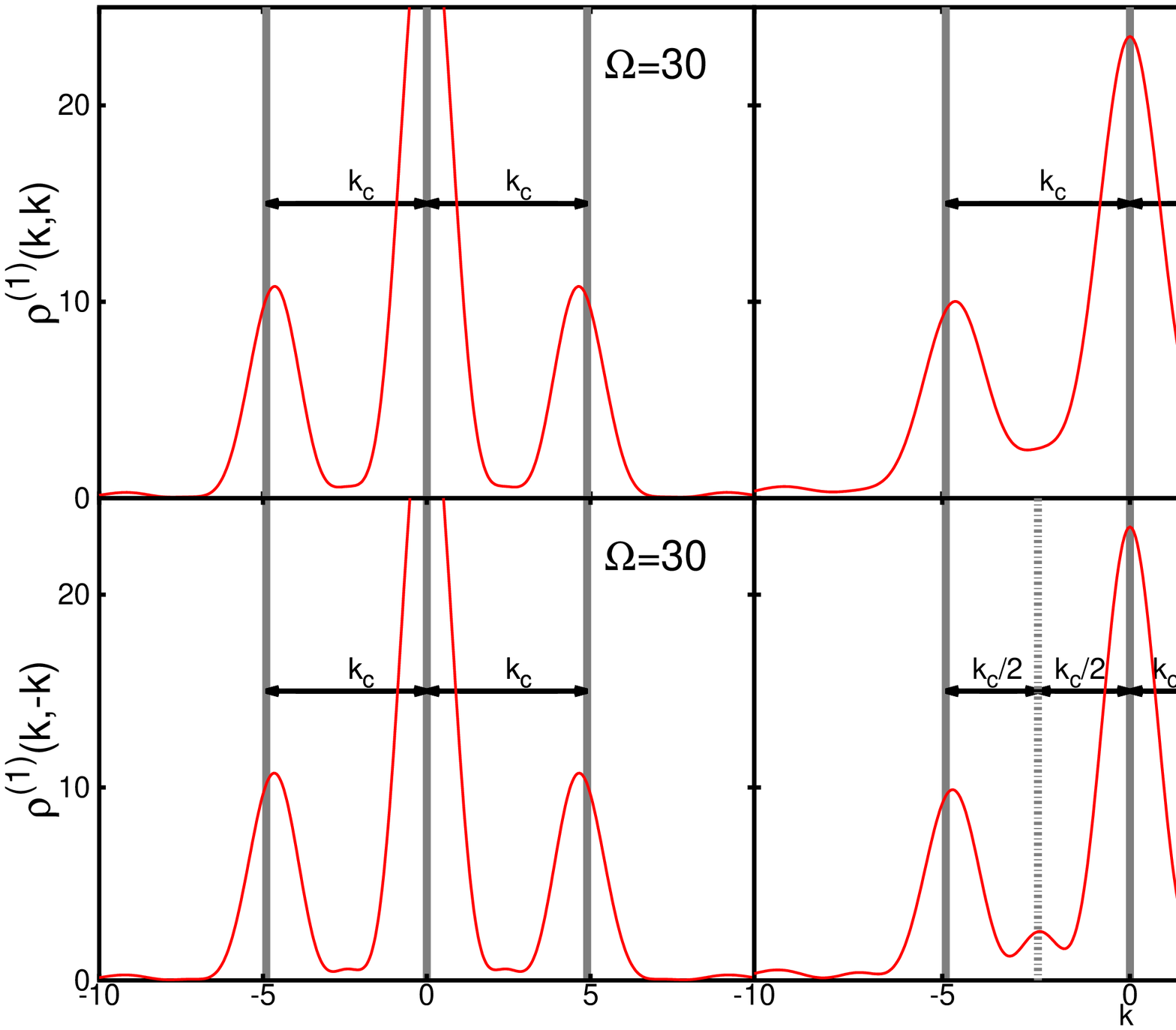}
 \caption{Emergence of fragmented superradiance and breakdown of the two-level description. The diagonal $\rho^{(1)}(k,k)$ (top) and skew-diagonal $\rho^{(1)}(k,-k)$ (bottom) of the reduced density matrix in momentum space are shown in the superradiant phase for pump powers $\Omega=30,40,100$ in the left, middle, and right column, respectively. In the superradiant but non-fragmented region, two side-peaks are visible at $\pm k_c$, where $k_c=4.9$ is the quasimomentum of the periodic potential spontaneously generated by the cavity photons (top and bottom left). This is as expected from the two-level picture. The side-peaks in the density $\rho^{(1)}(k,k)$ are gradually transformed to a broad single-maximum momentum distribution (top middle and right) indicating the breakdown of the two-level description in momentum space. The skew-diagonal $\rho^{(1)}(k,-k)$ shows the transition to fragmentation; the side-peak structure is transformed from a $\pm k_c$ 
to a $\pm k_c/2$ spacing. All quantities shown are dimensionless, see text for further discussion.}
 \label{Fig:2level}
\end{figure}

Here, we consider the validity of the description of the superradiant state with a two-level picture in momentum space (2LP).
We find that the system can be described with the 2LP when it is superradiant and not fragmented. When fragmentation emerges and the system enters the fragmented superradiant phase, however, the 2LP breaks down. To illustrate this behavior, we plot the momentum density $\rho^{(1)}(k,k)$ and the skew-diagonal of the reduced density matrix in momentum space, $\rho^{(1)}(k,-k)$ as a function of the pump power $\Omega$ in Fig.~\ref{Fig:2level}.
With the transition from superradiance to fragmented superradiance, the peaks at $\pm \hbar k_c$ that are the characteristic feature of the momenum distribution $\rho^{(1)}(k,k)$ in the 2LP disappear gradually. This shows the breakdown of the two-level picture for the fragmented superradiant state. In the skew-diagonal, however, a side-peak structure with a $\pm \hbar k_c/2$ spacing starts to form at the transition to fragmented superradiance. This structure maybe understood as an interference phenomenon: bosons that sit in different orbitals of the fragmented Bose-Einstein condensate couple \textit{independently} to the field in the cavity and the respective recoil momenta are superimposed.